# Multi-Channel Access Solutions for 5G New Radio

Nurul Huda Mahmood[1], Daniela Laselva[2], David Palacios[3], Mustafa Emara[4], Miltiades C. Filippou[4], Dong Min Kim[5], Isabel de-la-Bandera[3]

[1]*Center for Wireless Communications*
*University of Oulu*
Oulu, Finland
nurulhuda.mahmood@oulu.fi

[2]*Nokia Bell Labs*
Aalborg, Denmark
daniela.laselva@nokia-bell-labs.com

[3]*Communications Engineering Dept.*
*University of Malaga*
Malaga, Spain
{dpc, ibanderac}@ic.uma.es

[4]*Intel Deutschland GmbH*
Neubiberg, Germany
{mustafa.emara, miltiadis.filippou}@intel.com

[5]*Department of IoT*
*Soonchunhyang University*
Asan, South Korea
dmk@sch.ac.kr

*Abstract* – Multi-channel access is a family of multi-service radio resource management solutions that enable a user equipment to aggregate radio resources from multiple sources. The objective is multi-fold: throughput enhancement through access to a larger bandwidth tailored for enhanced mobile broadband, reliability improvement by increasing the diversity order and/or coordinated transmission/reception tailored for ultra-reliable low latency communication service classes, as well as flexibility and load balancing improvement by decoupling the downlink and the uplink access points, for both service classes. This paper presents several multi-channel access solutions for 5G New Radio multi-service scenarios. In particular, throughput enhancement and latency reduction concepts like multi-connectivity, carrier aggregation, downlink-uplink decoupled access and coordinated multi-point connectivity are discussed. Moreover, novel design solutions exploiting these concepts are proposed. Numerical evaluation of the introduced solutions indicates significant performance gains over state-of-the-art schemes; for example, our proposed component carrier selection mechanism leads to a median throughput gain of up to 100% by means of an implicit load balance. Therefore, the proposed multi-channel access solutions have the potential to be key multi-service enablers for 5G New Radio.

## I. Introduction

The Fifth-Generation New Radio (5G NR) is conceived to entail the ever-growing demand of multi-service mobile communications [1]. Compared to existing Long Term Evolution (LTE) networks, 5G NR requirements have expanded both vertically and horizontally. In the vertical domain, higher peak and average data rates are demanded for traditional mobile broadband services, corresponding to enhanced mobile broadband (eMBB) services, which are expected to allow achieving peak data rates of 20 Gbps. Horizontally, new service classes, such as ultra-reliable low-latency communications (URLLC), are introduced. Under URLLC, milli-second order latencies are expected, as well as extremely high reliability values (e.g., $1 - 10^{-5}$) [2].

Solutions proposed to meet the demanding performance requirements of eMBB and URLLC services span from physical layer (PHY) approaches, like the flexible numerology provided by 5G NR in terms of subcarrier spacing and the recently proposed short Transmission Time Interval (TTI) (specially devised for low-latency requirements) to Medium Access Control (MAC) layer solutions [3, 4], which allow the use of a flexible frame structure and the adoption of preemptive resource allocation for low-latency services. In addition, higher-layer solutions have also risen, like the packet duplication at the Packet Data Convergence Protocol (PDCP) level, providing further enhancements in terms of reliability [5].

Multi-channel access (MCA) is a promising family of radio resource management (RRM) solutions being explored towards this end. MCA allows a user equipment (UE) to simultaneously access multiple channels through one or more transmitting nodes. Carrier aggregation (CA) appears as an example of single-node MCA, whereas examples of multi-node MCA include coordinated multi-point (CoMP) access, multi-connectivity (MC) and downlink–uplink decoupling (DUDe) [6 - 9].

The aggregation of radio resources via MCA allows enhancing either the throughput, by splitting the data flow, or the reliability through data duplication. By splitting the data flow among different radio resources of different cells, a UE has access to a larger aggregated bandwidth, resulting in a throughput enhancement tailored for eMBB. Conversely, the duplication of the data flow through different cells allows the reception of multiple copies of the same data, thereby improving the reliability through diversity and/or repetition, for the benefit of URLLC. Thanks to their flexibility, MCA solutions are a key enabler to support multi-service scenarios by satisfying the various services' diverse performance requirements. However, the enhanced throughput and improved reliability with MCA are obtained at the expense of increased resource use and signaling load. Thus, optimization of MCA operations is necessary to reap its advantages, while minimizing the costs.

This article discusses MCA in 5G NR from the standpoint of ongoing 5G NR standardization activities within the 3rd Generation Partnership Project (3GPP), describes some of its key challenges in Section II, and presents several novel and promising MCA enhancements that address the identified challenges in Section III. An overview of the proposed solutions and the efforts needed for their standardization in 3GPP is summarized at the end.

## II. MULTI-CHANNEL ACCESS: TECHNOLOGY COMPONENTS AND CHALLENGES

### A. Multi-Connectivity

*Description & advantages*
MC is an extension of the dual connectivity (DC) functionality, where a UE can simultaneously use resources of multiple transmitting nodes. DC was first introduced in LTE as a throughput enhancement feature [7]. 3GPP has generalized the LTE DC design in 5G NR to enable the support of Multi-RAT DC (MR-DC), i.e., DC between 5G NR and LTE [6]. Reliability-oriented DC, where a given packet is independently transmitted through two different nodes, is also introduced in 5G [10]. The goal here is to increase the likelihood of successful reception by introducing radio link diversity. Henceforth, we discuss the general case of MC, which includes DC as a specific realization.

*Challenges*
In reliability-oriented MC, when focusing on the downlink scenario, a packet arriving at the PDCP-anchor node (known as the master node or MN) is duplicated at the PDCP layer and forwarded to the duplicating node(s) – known as secondary node(s) or SN – over the Xn interface, i.e., the same data packet is transmitted to the same UE through multiple links independently. The UE keeps the first successfully received packet, while it discards any subsequent copies of it. The Hybrid Automatic Repeat Request (HARQ) feedback (acknowledgment – ACK, or negative acknowledgment – NACK) for each transmission is fed back to the respective transmitting node. A NACK triggers a HARQ retransmission, even when a copy of the PDCP packet is successfully received through another node, resulting in unnecessary retransmissions. To overcome this limitation and increase resource efficiency, subsequent (re)transmissions of the same packet from other nodes should be avoided as soon as the packet is successfully received at the destination. It is noted that resource efficiency becomes a crucial challenge, especially in the existence of a large number of duplicates and/or nodes.

### B. Carrier Aggregation

*Description & advantages*
CA considers MCA from a bandwidth perspective and allows the aggregation of channels from different bands. In multi-node CA, the UE is assigned the primary cell (PCell) at the MN and the primary secondary cell (PSCell) at the SN. Any additional component carrier (CC) assigned at either node hosts a secondary cell (SCell). The CA setup is based on the same RRM measurements design as MC, where the PCell determines the suitability of potential SCells based on the UE reporting of Reference Signal Received Power/Reference Signal Received Quality (RSRP/RSRQ) measurements. However, rather than at the PDCP layer, multiple carriers can be aggregated at the MAC layer in CA, which controls the multiplexing of data and its transmission on the available CCs. It is noteworthy that the introduced PDCP packet duplication mechanism is also supported in conjunction with CA, constrained by the need for the MAC layer to guarantee that two duplicated packets are not transmitted on the same CC, so that the duplication benefits are preserved.

*Challenges*
Carrier aggregation expands the accessible bandwidth for a given UE by allowing simultaneous transmissions across multiple CCs. However, the allocation of CCs among the serving nodes is not straightforward, and hence, a proper CC assignment should be made on a per-UE basis. For example, an efficient algorithm design for CC allocation that limits the signal processing complexity, is needed to apply load-balancing methods and to avoid redundant transmissions, while increasing the UE throughput. Moreover, to successfully implement any throughput-oriented MCA solution, accurate and up-to-date information broadcasting has to be carried out jointly among the multiple next-generation nodeBs (gNB), either through over-the-air interface or via the Xn interface. For example, the throughput performance of data split MC for downlink communication is foreseen to be sensitive to the quality, capacity and delay of the Xn interface.

### C. Downlink/Uplink Decoupling

*Description & advantages*
Another dimension of MCA refers to the capability of performing distinct UE-cell associations for the DL and UL transmission directions. Cell densification, a promising solution to meet the stringent throughput requirements of 5G NR (e.g., high data rates in hot spots), leads to an imbalance in the downlink/uplink traffic [8]. This requires revisiting the performance optimality of the conventional downlink-driven RSRP based UE-cell association rule. To this end, DUDe, where the UE has the flexibility to be served in the downlink and the uplink by schedulers located at two different nodes, promises various performance benefits, e.g., increased uplink data throughput [9, 11].

*Challenges*
Communication delay reduction, which is especially relevant for URLLC (i.e., vehicular and industry 4.0) applications, is a challenging task for MCA solutions. At the same time, a limitation of heterogeneous networks (HetNets), is the large load imbalance among multiple tiers stemming from differences in inter-tier resource availability. To remedy such issues, MCA can be complemented by multi-access edge computing (MEC) technology, which provides an open, low-delay packet processing environment, in close proximity to end users [13]. However, focusing on HetNets and considering latency-intolerant applications, where the UEs have the option to delegate the execution of demanding tasks at a MEC server (e.g., physically co-located with a gNB), the application of conventional cell association is non-optimal, and a computationally-aware, DUDe-based UE-cell association rule needs to be applied, instead, as will be elaborated in Section III.C of the paper.

### D. Coordinated Multi-Point Access

*Description & advantages*
CoMP comprises multiple techniques that exploit different cooperation strategies among cells (or transmission points). Relying on fast network interfaces to exchange cooperation information, CoMP entails joint transmission (JT), i.e., the simultaneous transmission and reception of the same data to a UE via multiple nodes applying joint processing and reception at the lower layers (PHY and MAC), thus improving reliability. Alternatively, CoMP could also restrict the transmission to a UE to one cell at a time, while leveraging coordination among the cells to increase the signal to interference plus noise ratio, or control the interference at the scheduled transmission. This includes techniques such as coordinated scheduling, coordinated beamforming, and dynamic point selection.

*Challenges*

In relevance to the above explained challenge on the experienced communication delay, CoMP solutions devising packet duplication may prove challenging when coordination among the various transmitting nodes is performed by utilizing inter-node interfaces of low capacity and, possibly increased delay. Thus, a trade-off between coordination efficiency (e.g., in the sense of interference mitigation effectiveness) and packet transmission timeliness needs to be addressed for CoMP access, considering non-ideal fronthaul connections.

### III. MULTI-CHANNEL ACCESS SOLUTIONS AND PERFORMANCE EVALUATION

This section presents four promising solutions, corresponding to MC, CA, DUDe and CoMP access respectively, addressing the challenges highlighted in the previous section.

#### A. Novel Duplication Status Report for Multi-Connectivity

In this section, we propose a network discard mechanism that relies on a novel UE duplication status report. This report indicates to other node(s) in the duplication set that a certain PDCP packet has been successfully received, thus allowing the duplicating nodes to timely discard the flagged PDCP packet if it has not yet been transmitted.

Two key pieces of information are required at the UE to send the proposed duplication status report, namely, the reporting time and the reporting destination. When data duplication is activated for a UE, the corresponding duplication set is semi-statically configured, such that the UE can expect to receive duplicated PDCP packets only from these selected nodes. Therefore, the UE can target the duplication status report to the nodes in the duplication set. To reduce the reporting overhead, we propose to include a one-bit flag in the scheduling grant associated to a PDCP packet indicating that the packet in this physical transmission is scheduled for duplication from nodes in the duplication set.

Once the UE successfully decodes a flagged packet, it sends a conventional PHY ACK signal to the transmitting node, along with a duplication status report consisting of the packet identifier (the PDCP sequence number) to the other nodes in the duplication set. Upon receiving such a duplication status report, a node becomes aware that this particular packet has been successfully received at the UE, and thus, it does not need to (re)transmit it. As a result, the node will be able to discard this packet from its individual buffers (if not already transmitted), thereby reducing redundant transmissions. Fig. 1 depicts the proposed discard mechanism and compares it with the state-of-the-art, where duplicated PDCP packets are discarded at the UE.

In order to demonstrate its viability, the proposed discard mechanism is evaluated via Monte-Carlo simulations focusing on the performance of a specific URLLC user at the cell edge. In line with the ongoing 3GPP Release-16 study [6], which targets, among others, to increase the resource efficiency of duplication, the evaluation is limited to MC with two nodes, i.e., DC. A HetNet is considered with a single macro cell and a single small cell, operating at different frequency layers. The inter-frequency MC model is, thus, assumed. For further details, please refer to [10]. The macro cell is the serving cell in single-connectivity mode, while both the macro and small cell serve the UE in the MC mode. The mean signal-to-noise ratio (SNR) from both links is fixed at 10 dB, while the target SNR is set to 0 dB.

Fig. 2 presents the latency at $10^{-5}$ outage probability, and the transmission efficiency measured in terms of the average number of transmissions per delivered packets, for baseline single connectivity (SC), MC, and MC with the proposed discard feature. Under the considered setting, PDCP packet duplication via DC results in more than 50% latency reduction, as compared to the baseline single connectivity case at $10^{-5}$ outage probability. However, this is achieved at the expense of the resource efficiency, which is halved compared to SC. MC with the proposed discard feature achieves the same latency at $10^{-5}$ outage probability, as compared to conventional MC, alongside a transmission efficiency improvement of around 5%. Such gains result from the reduction in the number of duplicate transmissions, which, in turn, leads to lower queuing delays, a critical factor for multi-UE scenarios. Further optimization of the MC is the subject of our ongoing research.

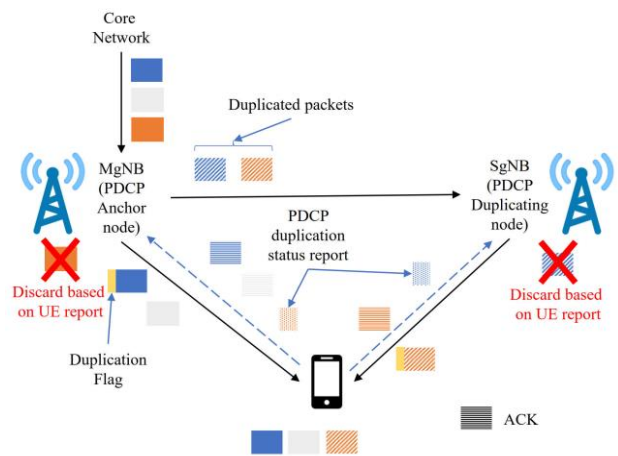

Fig. 1 – Implementation schematic of the proposed solution (bottom) compared to the state of the art (top).

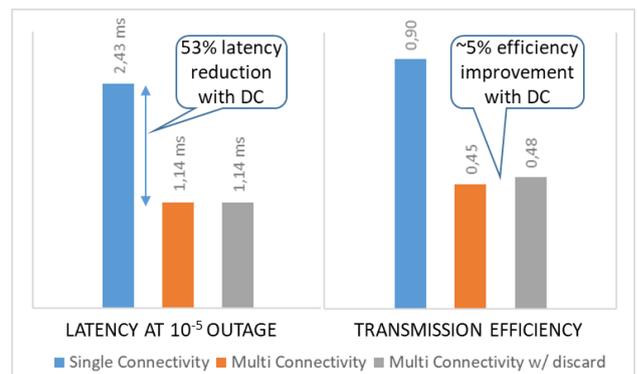

Fig. 2 – Performance Evaluation of MC with the proposed 'Duplication Status Report'.

#### B. Component Carrier Selection Mechanism for MCA

This solution addresses the automatic allocation of CCs to a UE in a generic manner, encompassing both single-node and multi-node MCA and relies on a rule-based system, which has been shown as a useful tool for optimization in the field of mobile communications [12]. Although it is described in the light of maximizing the throughput of eMBB UEs, for which the data flow is split among the assigned CCs, it is also applicable to URLLC services, where the data flow would be duplicated for reliability purposes.

Focusing on the proposed mechanism, the system aims at determining the number and indices of CCs to be assigned to a specific UE, as well as the gNB(s) providing them, according to the optimization objective(s). The antecedents of the rules are made up over performance information, gathered from the UEs (e.g., RSRP or RSRQ) and the CCs themselves (e.g., load information). The consequence of the rules are scores (standing for their suitability given a certain policy), over which an aggregation method is applied. Finally, the CCs with the highest aggregated scores above a minimum threshold are assigned to the users.

A proof of concept was implemented in a load-balancing environment. That is, a situation in which the heterogeneous spatial distribution of the users throughout the deployment area makes a reduced group of nodes support most of the offered traffic, leading to a high rate of service blockage, whereas many other nodes remain almost unused. To that end, the UE-reported RSRQ and the load level of a CC, derived from the instantaneous number of UEs allocated to a CC, have been used as input performance metrics. Different rules have been defined that assign low scores to low values of RSRQ and high-load levels, and vice versa. The final score of a CC is computed as the average of the scores provided by each rule.

The proposed solution is tested using a system-level simulator in a macro-cell scenario consisting of 12 tri-sector sites. Each sector is composed of five 1.4 MHz co-located CCs. A geographical region of a high user density is simulated to produce the load-imbalanced scenario. Two different sets of rules are evaluated: first, as a baseline, a case in which the rules only consider RSRP as input following the traditional approach for UE-gNB association, and second, the case in which both RSRQ and CC load metrics are assessed.

Fig. 3 shows the $5^{th}$, $50^{th}$ and $95^{th}$ percentiles of the UE throughput for the baseline case with dashed lines and the ones of the proposed solution with solid lines. This figure illustrates the achieved throughput levels with the number of CCs assigned to a UE ranging from one to five, depending on how many CCs scored above the minimum defined threshold. Evidently, the proposed CC selection method results in a significant UE throughput improvement resulting from efficient load balancing. More specifically, the proposed solution provides 66% average gain for the users experiencing the worst throughput values ($5^{th}$ percentile) over the state-of-the-art RSRP-based solution, and up to 75% gain at the peak throughput ($95^{th}$ percentile). As expected, higher numbers of CCs generally result in higher UE throughputs. In cases where more than one CC is assigned to a UE, each CC may be provided by a different node. For example, for the case of two CCs, approximately half of the UEs used CCs belonging to the same node (thus, applying CA), whereas the other half used CCs provided by different nodes (using DC). In multi-service scenarios, it is noted that the increased throughput shown above for eMBB can be converted in additional capacity for URLLC traffic.

*C. Latency-Reducing Connectivity based on Processing Proximity*

In this section, a new association metric is proposed for HetNets targeting latency reduction when UEs choose to offload demanding processing tasks to the network. Conventionally, considering a single tier of communication, the maximum downlink RSRP rule determines the cell to which the UE will be connected for both DL and UL

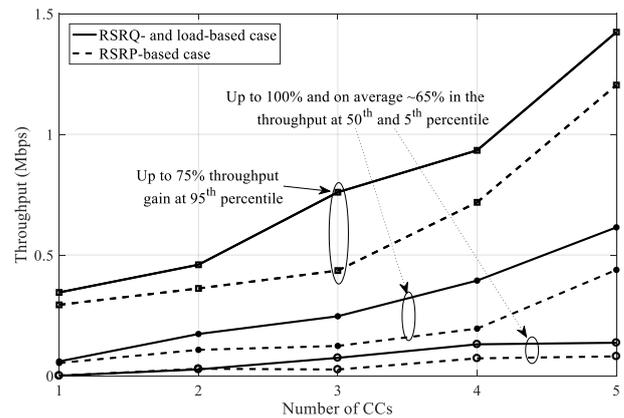

Fig. 3 – Per-UE achieved throughput for an increasing number of CCs; comparing the proposed solution for CC assignment with a traditional RSRP-based UE-gNB association

communication. Nevertheless, employing such a connectivity criterion in a highly heterogeneous deployment consisting of multi-tier nodes with diverse resource capabilities leads to load imbalance among the different tiers [8]. As a result, DUDe has been proposed as a disruptive solution for an enhanced network performance [9, 11], mainly giving the users the flexibility to associate with the gNB that provides the minimum pathloss, when it comes to UL communication.

Latency reduction through MEC technology stands out as an important feature with reference to URLLC, including use cases such as offloading a demanding processing task to the network. Hence, contrary to the state-of-the-art DUDe solution, which has been investigated in current technical literature, according to which, the UL association is conducted based on the minimum pathloss criterion, in our proposed solution, a UE, as a result of applying the proposed *computationally-aware* association rule for UL communication, will choose to connect to the closest gNB/MEC server with the largest available computational power.

In further detail, additional processing resources are granted by MEC servers, which are assumed physically collocated with the radio nodes. Under this assumption, with the aim of better balancing the user load and, thus, enhance latency reduction gains, the proposed solution is to apply different UE-cell association rules for DL and for UL communications. Focusing on UL communication, we propose an association rule that takes the available processing power offered by a MEC server into account, to best exploit the *computational proximity* to a given terminal, e.g., for the purpose of task offloading in the UL. Since the computational capacity of MEC servers is needed to form the proposed connectivity rule, it is assumed that such information is broadcast to the UEs by the multi-tier gNBs, as part of the system information block (SIB).

To quantify the performance gain of the proposed association metric, the Complementary Cumulative Distribution Function (CCDF) of the Extended-Packet Delay Budget (E-PDB) for a demanding offloaded task is depicted in Fig. 4, where E-PDB refers to the one-way latency composed of the radio connectivity latency in the uplink and the needed execution time at the selected MEC server. Focusing on a HetNet composed of two tiers, the E-PDB levels for the conventional (coupled, maximum RSRP-based) and the proposed, MEC-aware association rule for UL communication (hence, decoupled association for DL/UL) are shown. It should be noted that the parameter ω, termed as the *inter-tier cross-domain resource disparity*, is defined as the ratio between the radio resource disparity (i.e., transmit power) over the computational resource disparity (i.e., processing power) of the two involved network tiers. For more details regarding the system model and the system parameter values, the interested reader is referred to [13].

One can observe that the application of the proposed, computational proximity-based, decoupled association scheme results into a lower probability of violating a given E-PDB threshold with nearly 40% E-PDB reduction for the $50^{th}$-percentile of UEs. In other words, the experienced UL latency is decreased when employing the proposed cell association rule which facilitates DUDe. This occurs due to the enhanced load balance between the different tiers leading towards better exploitation of the available computational resources at the associated radio node. For different values of the parameter ω, an adaptive association procedure may be considered to minimize the experienced E-PDB when a UE decides to offload a demanding task to the network.

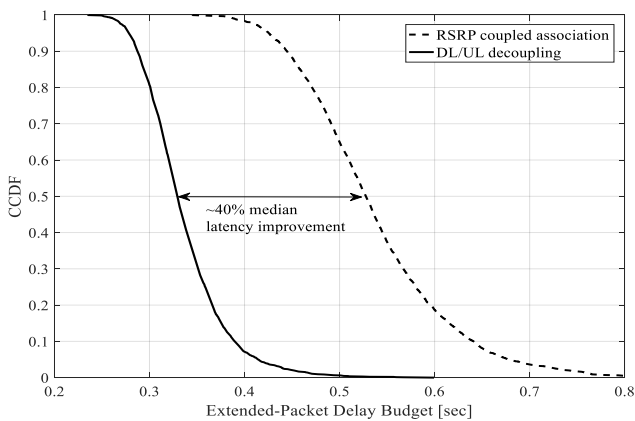

Fig. 4 – E-PDB CCDF comparing the (conventional) coupled and the proposed (MEC-aware) decoupled association rules for ω = 2 and an average processing workload of 1000 CPU cycles/ input bit.

### D. Coordinated Multi-Point Connectivity for Low-Latency Applications

The 5G NR protocol must work properly for smooth data transmission, and control information must be delivered in a timely manner. For example, the transmitted data in one direction will be acknowledged with the opposite direction response. Such information exchange forms a two-way communication. If there is a very large number of users, i.e., a ultra-dense network (UDN), in the existence of mixed uplink and downlink traffic, it is important to provide a fast control message response, so that the sender can decide on the subsequent action immediately. To achieve this, we can adopt MC and CoMP.

We assume that users are divided into two groups: low latency users (LLUs) and latency tolerant users (LTUs). Consider two neighboring gNBs that are connected through the Xn interface over fiber, as well as two users, one LLU and one LTU. These cooperating gNBs can serve their users jointly by exchanging user information. If the performance requirements of services are different, they must be handled differently. For example, if LLU and LTU coexist, LLU can be processed first since transmission of LTU can tolerate delays. However, if LLU and LLU coexist, one way is to serve these users at the same time by using a technique such as CoMP. Depending on the transmission direction, the following cooperation can be achieved.

*Cooperation with same directional traffic:* If both users have DL traffic as shown in Fig. 5, cooperating gNBs will schedule the LLU using reliability-oriented DC (left of Fig. 5, case a). If both users are LTUs, cooperating BSs schedule one user randomly via reliability-oriented DC so that the packet exits the buffer quickly (right of Fig. 5, case a). If both are LLU, then cooperating BSs schedule them jointly using non-coherent JT-CoMP.

*Cross-link cooperation:* If both users have cross directional traffic, the UL-gNB can exploit information provided from the DL-BS via Xn interface to execute Interference Cancellation (IC) to perform IC-CoMP [14]. It is assumed that UL and DL share the same frequency. However, the same technique can be applied to separate UL and DL frequencies. In this case, cross-directional traffic might be preferable because it is interference-free and will be easier to implement.

To quantify the performance, we consider a scenario where the users transmit their data and receive ACK/NACK. If a user does not receive an ACK, it retransmits the data. We assume that the UL and DL time slots have the same length, which is normalized to one. The two-way latency is defined as the number of consumed time slots from the moment of first data symbol transmission until the ACK is received. Simulation results show that the average two-way latency is reduced by 60% compared to the baseline single-connectivity scheme. Detailed results are presented in [15].

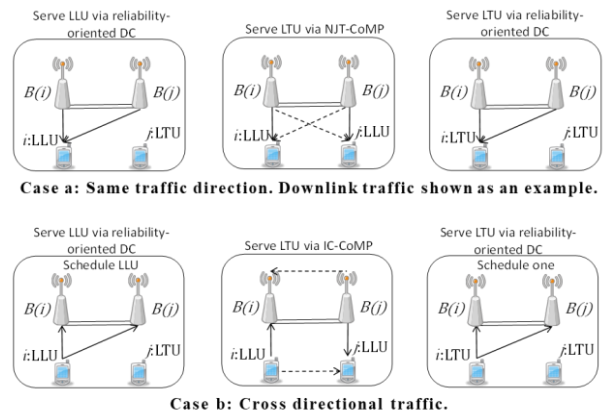

Fig. 3 – Both users have the same directional traffic (case a) or cross directional traffic (case b). Based on the type of the traffic, each user could be low-latency user (LLU) or latency-tolerant user (LTU).

### E. Overview of the Proposed MCA Solutions

An overview of the proposed MCA solutions is shown in Table I, along with the standardization effort needed to have the techniques supported in 3GPP 5G NR standards.

## IV. CONCLUSION

The emergence of new services and requirements for future 5G NR demands novel RRM techniques. This paper provides an overview of MCA solutions tailored to user plane enhancements of 5G NR comprising throughput boosting, reliability improvement and latency reduction. In particular, carrier aggregation, multi-connectivity in conjunction with data duplication, downlink-uplink decoupling and coordinated multi-point transmission schemes are presented, describing their current developments and anticipated evolution. Novel solutions addressing key challenges identified for each MCA concept are proposed and numerically validated. The standardization effort needed to implement these solutions in 5G NR is also outlined. All the proposed solutions are found to provide promising performance benefits over state-of-the-art solutions. Although the proposed schemes address different challenges, they are compatible and focused towards the same goal: enabling 5G NR requirements in a multi-service context.

TABLE I – OVERVIEW OF THE PROPOSED MCA SOLUTIONS

| Scheme | Addressed Challenge/Benefit | Targeted Traffic Type(s) | Standardization Effort | Performance gain over state of the art |
|---|---|---|---|---|
| **Duplication Status Report for MC** | Design of a novel PDCP duplication status report to acknowledge reception of PDCP PDUs to multiple nodes and corresponding discard | Increased URLLC reliability, minimized duplication costs (interference / queueing delays) | URLLC and traffic mixes which include URLLC traffic and eMBB traffic | Duplication flag UE sending a short ACK to nodes in the duplication set | - 53% latency gain<br>- 5% transmission efficiency gain |
| **Component Carrier Selection for MCA** | Design of a rule-based CC selection per user, according to UE- and cell-level information | Fulfilled network operators' policies. E.g., throughput increase, load balancing, etc | Demonstrated for eMBB, can also benefit URLLC | Network exchanges for collecting the CC scores to enable their comparison | - 65% average throughput gain<br>- Up to 75% peak throughput gain |
| **Latency-Reducing Connectivity based on Processing Proximity** | Design of a computationally-aware UE-connectivity framework aiming at latency minimization in HetNets | Exploiting DL/UL decoupled cell association | Mainly URLLC, can also benefit eMBB | Novel system information to indicate MEC processing capabilities in UE vicinity | - 40% median latency improvement |
| **Coordinated Multi-Point Connectivity for Low-Latency Applications** | Method to manage additional links for duplicated transmissions to enhance reliability and latency | Flexibly and cooperatively decode received data, to enhance reliability | URLLC and traffic mixes which include URLLC traffic and eMBB traffic | N/A | - Two-way latency reduced by 60% |

## Acknowledgements

This work has partly been performed in the framework of the Horizon 2020 project ONE-5G (ICT-760809) receiving funds from the European Union, and partly under Academy of Finland 6Genesis Flagship (grant no. 318927). The authors would like to acknowledge the contributions of their colleagues in the project, although the views expressed in this work are those of the authors and do not necessarily represent the project. The work was carried out while N.H. Mahmood was at Aalborg University, Denmark.